\documentclass[reprint,prl,superscriptaddress]{revtex4-1}
\usepackage{bm}

\newcommand\beq{\begin{equation}}
\newcommand\eeq{\end{equation}}
\newcommand\beqn{\begin{eqnarray}}
\newcommand\eeqn{\end{eqnarray}}

\newcommand\cl{C_{\ell}}

\newcommand\nn{\nonumber}

\usepackage{graphicx}
\usepackage[large]{subfigure}
\usepackage{amssymb, amsmath}
\usepackage[amssymb]{SIunits}
\usepackage{aas_macros}
\usepackage{natbib}
\renewcommand\section[1]{\emph{#1}.---}

\begin{document}

\title{Evidence for dark energy from the cosmic microwave background alone using the Atacama Cosmology Telescope lensing measurements}

\author{Blake~D.~Sherwin}\affiliation{Dept.~of Physics,
Princeton University, Princeton, NJ, USA 08544}
\author{Joanna~Dunkley}\affiliation{Dept.~of Astrophysics, Oxford University, Oxford, 
UK OX1 3RH}\affiliation{Dept.~of Physics,
Princeton University, Princeton, NJ, USA 08544}\affiliation{Dept.~of Astrophysical Sciences, Peyton Hall, Princeton University, Princeton, NJ USA 08544}
\author{Sudeep~Das}\affiliation{BCCP, Dept.~of Physics, University of California, Berkeley, CA, USA 94720}\affiliation{Dept.~of Physics,
Princeton University, Princeton, NJ, USA 08544}\affiliation{Dept.~of Astrophysical Sciences, Peyton Hall, Princeton University, Princeton, NJ USA 08544}
\author{John~W.~Appel}\affiliation{Dept.~of Physics,
Princeton University, Princeton, NJ, USA 08544}
\author{J.~Richard~Bond}\affiliation{CITA, University of
Toronto, Toronto, ON, Canada M5S 3H8}
\author{C.~Sofia~Carvalho}\affiliation{IPFN, IST, Av.~Rovisco Pais, 1049-001 Lisboa, Portugal}\affiliation{Academy of Athens, RCAAM, Soranou Efessiou, 11-527 Athens, Greece}
\author{Mark~J.~Devlin}\affiliation{Dept.~of Physics and Astronomy, University of
Pennsylvania, Philadelphia, PA, USA 19104}
\author{Rolando~D\"{u}nner}\affiliation{Departamento de Astronom{\'{i}}a y Astrof{\'{i}}sica, Pontific\'{i}a Univ. Cat\'{o}lica,
Casilla 306, Santiago 22, Chile}
\author{Thomas~Essinger-Hileman}\affiliation{Dept.~of Physics,
Princeton University, Princeton, NJ, USA 08544}
\author{Joseph~W.~Fowler}\affiliation{NIST Quantum Devices Group, 325
Broadway Mailcode 817.03, Boulder, CO, USA 80305}\affiliation{Dept.~of Physics,
Princeton University, Princeton, NJ, USA 08544}
\author{Amir~Hajian}\affiliation{CITA, University of
Toronto, Toronto, ON, Canada M5S 3H8}\affiliation{Dept.~of Astrophysical Sciences, Peyton Hall, 
Princeton University, Princeton, NJ USA 08544}\affiliation{Dept.~of Physics,
Princeton University, Princeton, NJ, USA 08544}
\author{Mark~Halpern}\affiliation{Dept.~of Physics and Astronomy, University of
British Columbia, Vancouver, BC, Canada V6T 1Z4}
\author{Matthew~Hasselfield}\affiliation{Dept.~of Physics and Astronomy, University of
British Columbia, Vancouver, BC, Canada V6T 1Z4}
\author{Adam~D.~Hincks}\affiliation{Dept.~of Physics,
Princeton University, Princeton, NJ, USA 08544}
\author{Ren\'ee~Hlozek}\affiliation{Dept.~of Astrophysics, Oxford University, Oxford, 
UK OX1 3RH}
\author{John~P.~Hughes}\affiliation{Dept.~of Physics and Astronomy, Rutgers, 
The State University of New Jersey, Piscataway, NJ USA 08854-8019}
\author{Kent~D.~Irwin}\affiliation{NIST Quantum Devices Group, 325
Broadway Mailcode 817.03, Boulder, CO, USA 80305}
\author{Jeff~Klein}\affiliation{Dept.~of Physics and Astronomy, University of
Pennsylvania, Philadelphia, PA, USA 19104}
\author{Arthur~Kosowsky}\affiliation{Dept.~of Physics and Astronomy, University of Pittsburgh, 
Pittsburgh, PA, USA 15260}
\author{Tobias~A.~Marriage}\affiliation{Dept.~of Physics and Astronomy, The Johns Hopkins University, Baltimore, MD 21218-2686}\affiliation{Dept.~of Astrophysical Sciences, Peyton Hall, 
Princeton University, Princeton, NJ USA 08544}
\author{Danica~Marsden}\affiliation{Dept.~of Physics and Astronomy, University of
Pennsylvania, Philadelphia, PA, USA 19104}
\author{Kavilan~Moodley}\affiliation{Astrophysics and Cosmology Research Unit, Univ. of KwaZulu-Natal, Durban, 4041,
South Africa}
\author{Felipe Menanteau}\affiliation{Dept.~of Physics and Astronomy, Rutgers, 
The State University of New Jersey, Piscataway, NJ USA 08854-8019}
\author{Michael~D.~Niemack}\affiliation{NIST Quantum Devices Group, 325
Broadway Mailcode 817.03, Boulder, CO, USA 80305}\affiliation{Dept.~of Physics,
Princeton University, Princeton, NJ, USA 08544}
\author{Michael~R.~Nolta}\affiliation{CITA, University of
Toronto, Toronto, ON, Canada M5S 3H8}
\author{Lyman~A.~Page}\affiliation{Dept.~of Physics,
Princeton University, Princeton, NJ, USA 08544}
\author{Lucas~Parker}\affiliation{Dept.~of Physics,
Princeton University, Princeton, NJ, USA 08544}
\author{Erik~D.~Reese}\affiliation{Dept.~of Physics and Astronomy, University of
Pennsylvania, Philadelphia, PA, USA 19104}
\author{Benjamin~L.~Schmitt}\affiliation{Dept.~of Physics and Astronomy, University of
Pennsylvania, Philadelphia, PA, USA 19104}
\author{Neelima~Sehgal}\affiliation{KIPAC, Stanford
University, Stanford, CA, USA 94305-4085}
\author{Jon~Sievers}\affiliation{CITA, University of
Toronto, Toronto, ON, Canada M5S 3H8}
\author{David~N.~Spergel}\affiliation{Dept.~of Astrophysical Sciences, Peyton Hall, 
Princeton University, Princeton, NJ USA 08544}
\author{Suzanne~T.~Staggs}\affiliation{Dept.~of Physics,
Princeton University, Princeton, NJ, USA 08544}
\author{Daniel~S.~Swetz}\affiliation{Dept.~of Physics and Astronomy, University of
Pennsylvania, Philadelphia, PA, USA 19104}\affiliation{NIST Quantum Devices Group, 325
Broadway Mailcode 817.03, Boulder, CO, USA 80305}
\author{Eric~R.~Switzer}\affiliation{Kavli Institute for Cosmological Physics, 
5620 South Ellis Ave., Chicago, IL, USA 60637}\affiliation{Dept.~of Physics,
Princeton University, Princeton, NJ, USA 08544}
\author{Robert~Thornton}\affiliation{Dept.~of Physics and Astronomy, University of
Pennsylvania, Philadelphia, PA, USA 19104}\affiliation{Dept.~of Physics , West Chester University 
of Pennsylvania, West Chester, PA, USA 19383}
\author{Katerina~Visnjic}\affiliation{Dept.~of Physics,
Princeton University, Princeton, NJ, USA 08544}
\author{Ed~Wollack}\affiliation{Code 553/665, NASA/Goddard Space Flight Center,
Greenbelt, MD, USA 20771}


\begin{abstract}
For the first time, measurements of the cosmic microwave background radiation (CMB) alone favor cosmologies with $w=-1$ dark energy over models without dark energy at a 3.2-sigma level. We demonstrate this by combining the CMB lensing deflection power spectrum from the Atacama Cosmology Telescope with temperature and polarization power spectra from the Wilkinson Microwave Anisotropy Probe. The lensing data break the geometric degeneracy of different cosmological models with similar CMB temperature power spectra. Our CMB-only measurement of the dark energy density $\Omega_\Lambda$ confirms other measurements from supernovae, galaxy clusters and baryon acoustic oscillations, and demonstrates the power of CMB lensing as a new cosmological tool.
\end{abstract}
\maketitle

\section{Introduction}
Observations made over the past two decades suggest a standard cosmological model for the contents and geometry of the universe, as well as for the initial fluctuations that seeded cosmic structure \citep{riess/etal:1998,perlmutter/etal:1999,spergel/etal:2003}. The data imply that our universe at the present epoch has a dominant stress-energy component with negative pressure, known as ``dark energy'', and has zero mean spatial curvature. The cosmic microwave background (CMB) has played a crucial role in constraining the fractional energy densities in matter, $\Omega_m$, and in dark energy or the cosmological constant, $\Omega_\Lambda$ (or equivalently in curvature $\Omega_K=1-\Omega_\Lambda-\Omega_m$) \citep[e.g.,][]{komatsu/etal:2011}. Throughout this letter, we restrict our analysis to the simplest dark energy models with equation of state parameter $w=-1$.

The existence of dark energy, first directly observed by supernova measurements \citep{riess/etal:1998,perlmutter/etal:1999}, is required \cite{spergel/etal:2003} by the combination of CMB power spectrum measurements and any {\it one} of the following low redshift observations  \citep{hicken/etal:2009,kessler/etal:2009,reid/etal:2010,riess/etal:2009,vikhlinin/etal:2009}: measurements of the Hubble constant, measurements of the galaxy power spectrum, galaxy cluster abundances, or supernova measurements of the redshift-distance relation. At present, the combination of low-redshift astronomical observations with CMB data can constrain cosmological parameters in a universe with both vacuum energy and curvature to better than a few percent \citep{komatsu/etal:2011}.

However, from the CMB alone, it has not been possible to convincingly demonstrate the existence of a dark energy component, or that the universe is geometrically flat \citep{spergel/etal:2003,komatsu/etal:2011}. This is due to the ``geometric degeneracy'' which prevents both the curvature and expansion rate from being determined simultaneously from the CMB alone \cite{doroshkevich/zeldovich/sunyaev:1978,bond/efstathiou/tegmark:1997,zaldarriaga/spergel/seljak:1997}. The degeneracy can be understood as follows. The first peak of the CMB temperature power spectrum measures the angular size of a known physical scale: the sound horizon at decoupling, when the CMB was last scattered by free electrons. However, very different cosmologies can project this sound horizon onto the same degree-scale angle on the sky: from a young universe with a large vacuum energy and negative spatial curvature, to the standard spatially flat cosmological model, to an old universe with no vacuum energy, positive spatial curvature, and a small Hubble constant \citep{dunkley/etal:2009}. These models, therefore, cannot be significantly distinguished using only primordial CMB power spectrum measurements.

By observing the CMB at higher resolution, however, one can break the geometric degeneracy using the effect of gravity on the CMB \citep{stompor/efstathiou:1999}: the deflection of CMB photons on arcminute scales due to gravitational lensing by large scale structure. This lensing of the CMB can be described by a deflection field $\mathbf{d}(\mathbf n)$ which relates the lensed and unlensed temperature fluctuations $\delta T,\delta \tilde T$ in a direction $\mathbf n$ as $\delta T(\mathbf{n}) = \delta \tilde T(\mathbf{n} + \mathbf{d})$. The lensing signal, first detected at $3.4\sigma$ from the cross-correlation of radio sources with WMAP data \cite{smith/zahn/dore:2007} and at $4\sigma$ from the CMB alone by the Atacama Cosmology Telescope (ACT) \cite{das/etal:prep}, is sensitive to both the growth of structure in recent epochs and the geometry of the universe \citep{lewis/challinor:2006}.  Combining the low-redshift information from CMB
lensing with CMB power spectrum data gives significant constraints on $\Omega_\Lambda$, which 
the power spectrum alone is unable to provide.

\begin{figure}
\includegraphics[width=1.\columnwidth]{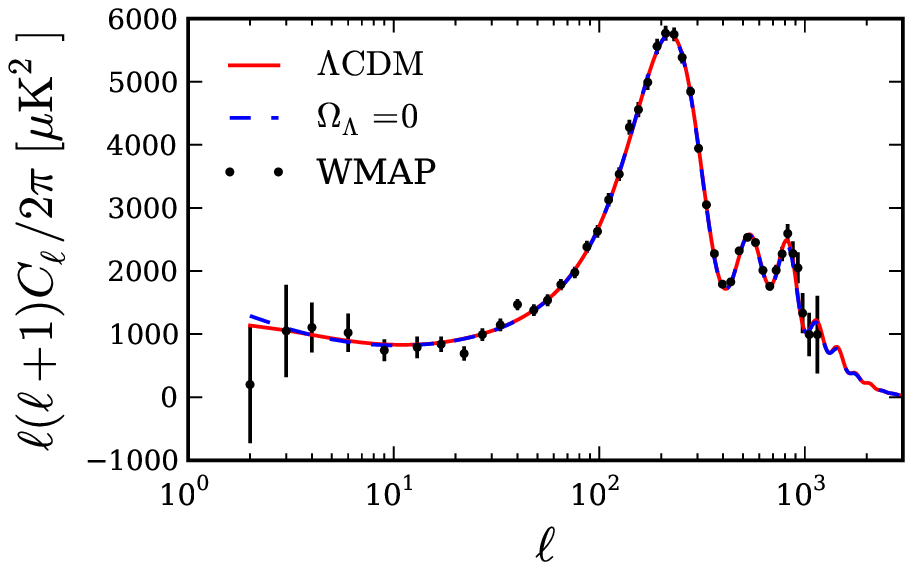}
\includegraphics[width=1.\columnwidth]{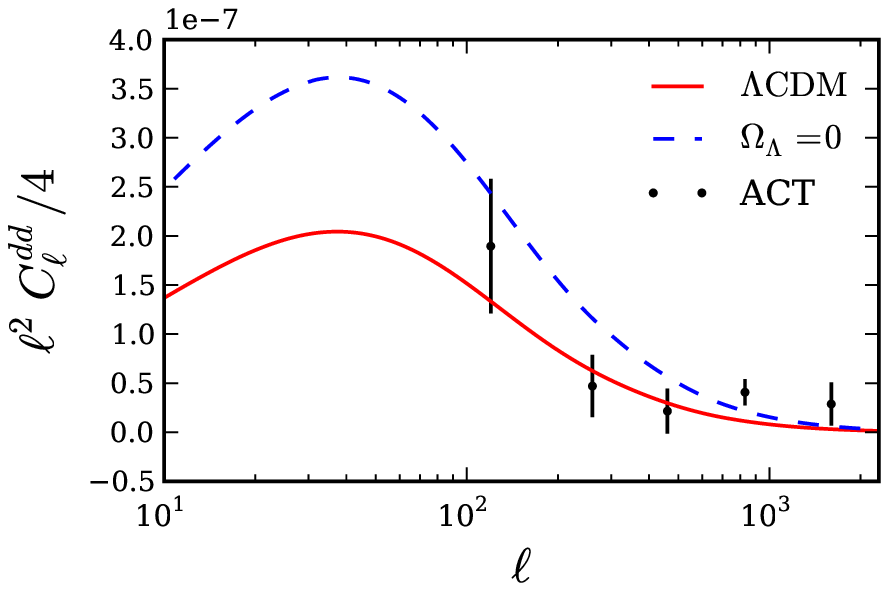}
\caption{Upper panel: Angular power spectra of CMB temperature fluctuations for two geometrically degenerate cosmological models, one the best-fit curved universe with no vacuum energy ($\Omega_\Lambda=0, \Omega_m=1.29$), and one the best-fit flat $\Lambda$CDM model with $\Omega_\Lambda=0.73, \Omega_m=0.27$. The seven-year WMAP temperature power spectrum data \cite{larson/etal:2010} are also shown; they do not significantly favor either model. \\
Lower panel: The CMB lensing deflection power spectra are shown for the same two models. They are no longer degenerate: the $\Omega_\Lambda=0$ universe would produce a lensing power spectrum larger than that measured by ACT \cite{das/etal:prep} (shown above).
\label{fig:temp}}
\end{figure}

The constraining power of the CMB lensing measurements is apparent in a comparison between two models consistent with the CMB temperature power spectrum (see Fig.~\ref{fig:temp}): the spatially flat $\Lambda \mathrm{CDM}$ model with dark energy which best fits the WMAP seven-year data \cite{larson/etal:2010} and a model with positive spatial curvature but without dark energy.
The two theory temperature spectra (and the temperature-polarization cross-correlation spectra) differ only at the largest scales with multipoles $\ell<10$, where the cosmic variance errors are large.
(The differences are due to the Integrated Sachs-Wolfe (ISW) effect, a large-scale CMB distortion induced by decaying gravitational potentials in the presence of dark energy \citep{boughn/crittenden/turok:1998} or, with the opposite sign, induced by growing potentials in the presence of positive curvature.) The polarization power spectra in the two models are also very similar.

However, these two cosmologies predict significantly different CMB lensing deflection power spectra $\cl^{dd}$. Fig.~\ref{fig:temp} shows that the universe with $\Omega_\Lambda=0$ produces more lensing on all scales. The ACT measurements shown in Fig.~\ref{fig:temp} are a better fit to the model with vacuum energy than to the model without dark energy.

Why is the lensing power spectrum higher in a universe without dark energy but with the same primordial CMB spectrum? This can be understood from the expression for the power spectrum of lensing deflection angles \citep{lewis/challinor:2006}:
\beqn
\label{eq:estimator}
 \frac{\ell^2}{4} {\cl}^{d d} &=  &   \int_0^{\eta_*} d \eta  \underbrace{   W^2(\eta)  }_{\mathrm{geometry}}~  \underbrace{\left[ D\left(\eta \right)/a(\eta)  \right]^2}_{\mathrm{growth}}
\eeqn
where $\eta$ is conformal lookback distance, $\eta_*$ is the conformal distance to the CMB last scattering surface, $D$ is the growth factor of matter perturbations since decoupling, $a$ is the scale factor, and $W(\eta)$ is a geometry and projection term given by
\beqn
\nn W(\eta) = \frac 32 \Omega_m H_0^2 ~ \frac{d_A(\eta_*-\eta)}{d_A(\eta_*)} ~P^{1/2}\left(k = \frac{\ell + 1/2}{d_A(\eta)},\eta_*\right).
\eeqn
where $H_0$ is the Hubble constant, $d_A$ is comoving angular diameter distance, $P(k,\eta_*)$ is the matter power spectrum at decoupling and $k$ is the comoving wavenumber.

\begin{figure} [h!]

\includegraphics[width=1.\columnwidth]{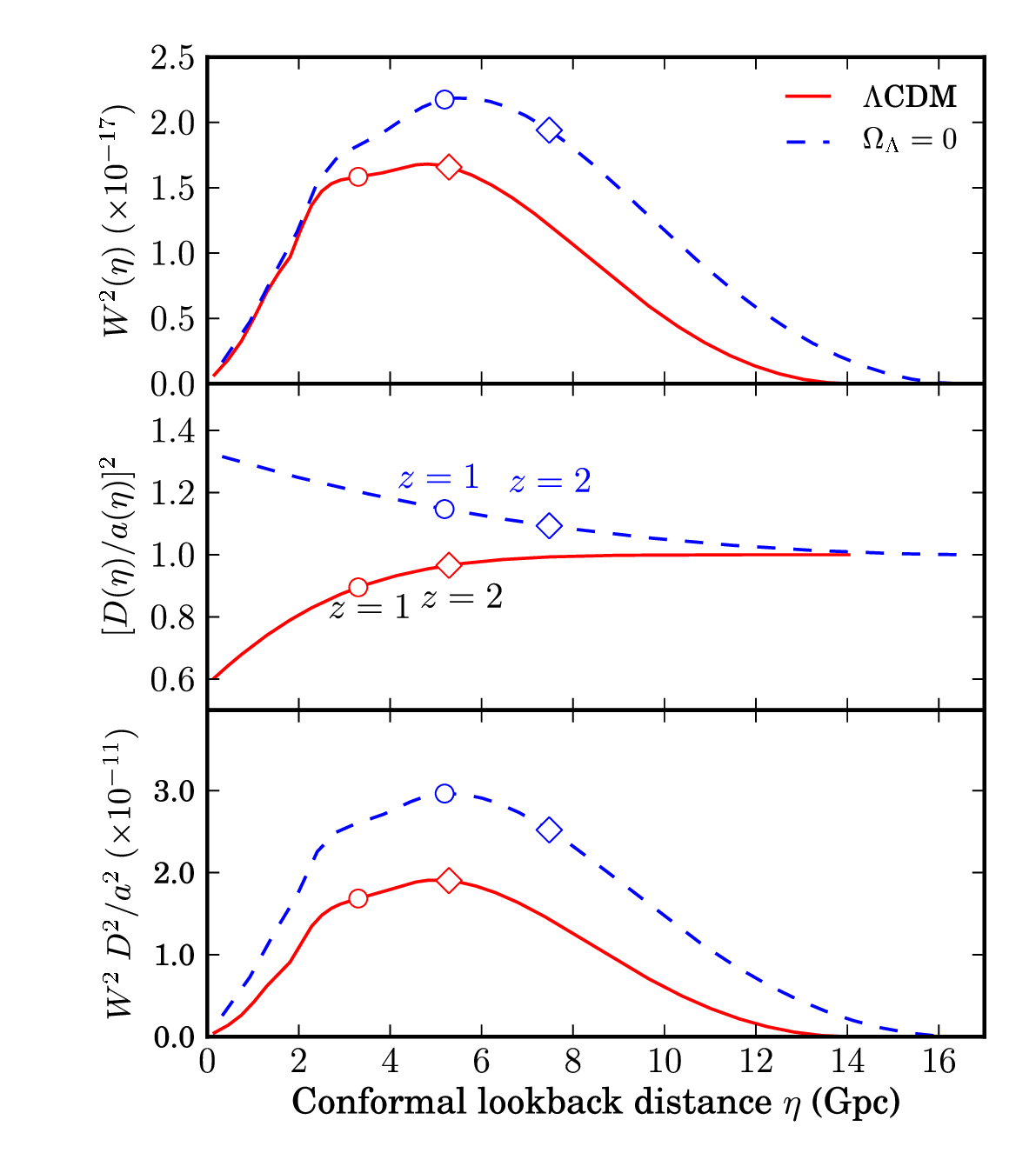}
\caption{Different terms in the kernel of the lensing integral of Eq.~\ref{eq:estimator} as a function of conformal lookback distance for $\ell=120$, for models as in Fig.~\ref{fig:temp}. Upper panel: geometry term. Middle panel: growth term, scaled to its value at decoupling for clarity. Bottom panel: total kernel.
\label{fig:kernel}}
\end{figure}

A plot of the kernel of the lensing integral in this equation, as well as its constituent ``geometry'' and ``growth'' terms, is shown in Fig.~\ref{fig:kernel} for both $\Lambda \mathrm{CDM}$ and $\Omega_\Lambda=0$ models. This figure shows that increased lensing in universes without dark energy is due to three effects: (1) CMB photons in a universe without dark energy spend more time at lower redshifts where structure is larger; (2) structure and potentials grow more in a universe with 
$\Omega_\Lambda=0$ and positive curvature; (3) in a universe without dark energy, projection effects pick out longer wavelength fluctuations which are larger for most lensing scales.

As the amplitude of the lensing signal is sensitive to $z < 5$ physics, measurements of CMB lensing break the geometric degeneracy and improve constraints on cosmological parameters.   In this letter, we construct a likelihood function by combining ACT lensing measurements and WMAP power spectra, and explore the new CMB-only parameter constraints resulting from the inclusion of lensing data.

\section{Methodology}
We fit a joint distribution for a set of cosmological parameters $\theta$ to our data $D$ (see, e.g. \citep{larson/etal:2010}).  In our analysis, we consider the following cosmological parameters:
\begin{equation}
\mathbf{\theta} = \{ \Omega_\Lambda, \Omega_K, \Omega_b h^2, \Omega_c h^2, n_s, \Delta^2_\mathcal{R}, \tau, A_{\mathrm{SZ}}\}
\end{equation}
where $\Omega_b h^2$ is the baryon density, $\Omega_c h^2$ is the cold dark matter (CDM) density, $n_s$ is the spectral tilt of the density fluctuations, $\Delta^2_\mathcal{R}$ is their amplitude (defined at pivot scale $k_0 = 0.002/\mathrm{Mpc}$), $\tau$ is the optical depth to reionization, and $A_{\mathrm{SZ}}$ is the amplitude of the WMAP V-band SZ template \citep{komatsu/seljak:2002}. The Hubble constant, $H_0\equiv100~h$~km/s/Mpc, can be derived from these parameters. The estimated distribution is the product of the likelihood $p(D| {\cl} (\mathbf{\theta}))$ and the prior $p(\theta)$.  Here $\cl(\theta)$ is the set of theoretical angular power spectra (CMB temperature power spectrum $\cl^{TT}$, CMB polarization power spectra $\cl^{TE}$ and $\cl^{EE}$, and lensing deflection angle power spectrum $\cl^{dd}$) derived from the parameters $\theta$. Uniform priors are placed on all sampled parameters. We use data from the WMAP seven-year temperature and polarization observations \citep{larson/etal:2010}, which map the CMB anisotropy over the full sky.  These are combined with the ACT lensing deflection power spectrum described in \citep{das/etal:prep}, obtained from a measurement of the lensing non-Gaussianity in a 324 deg$^2$ patch of the ACT equatorial CMB maps. The data were found to be effectively free of contamination from astrophysical sources or noise, with errors that were estimated to be Gaussian and uncorrelated. Since the correlation between the datasets is negligible, the likelihood is the product of the WMAP likelihood, $p(D_{\rm WMAP}| {\cl}^{\rm TT,TE,EE}(\mathbf{\theta}))$, described in \citep{larson/etal:2010}, and the ACT lensing likelihood, $p(D_{\rm ACT}| {\cl}^{d d}(\mathbf{\theta}))$ \cite{das/etal:prep}.

Theoretical CMB temperature and lensing power spectra are computed using the CAMB code \cite{lewis/challinor/lasenby:2000}. We follow the same approach as \cite{dunkley/etal:2009,larson/etal:2010}  to map out the posterior distribution of the parameters.

\section{Results}
\begin{figure} [h!]
\includegraphics[width=1.\columnwidth]{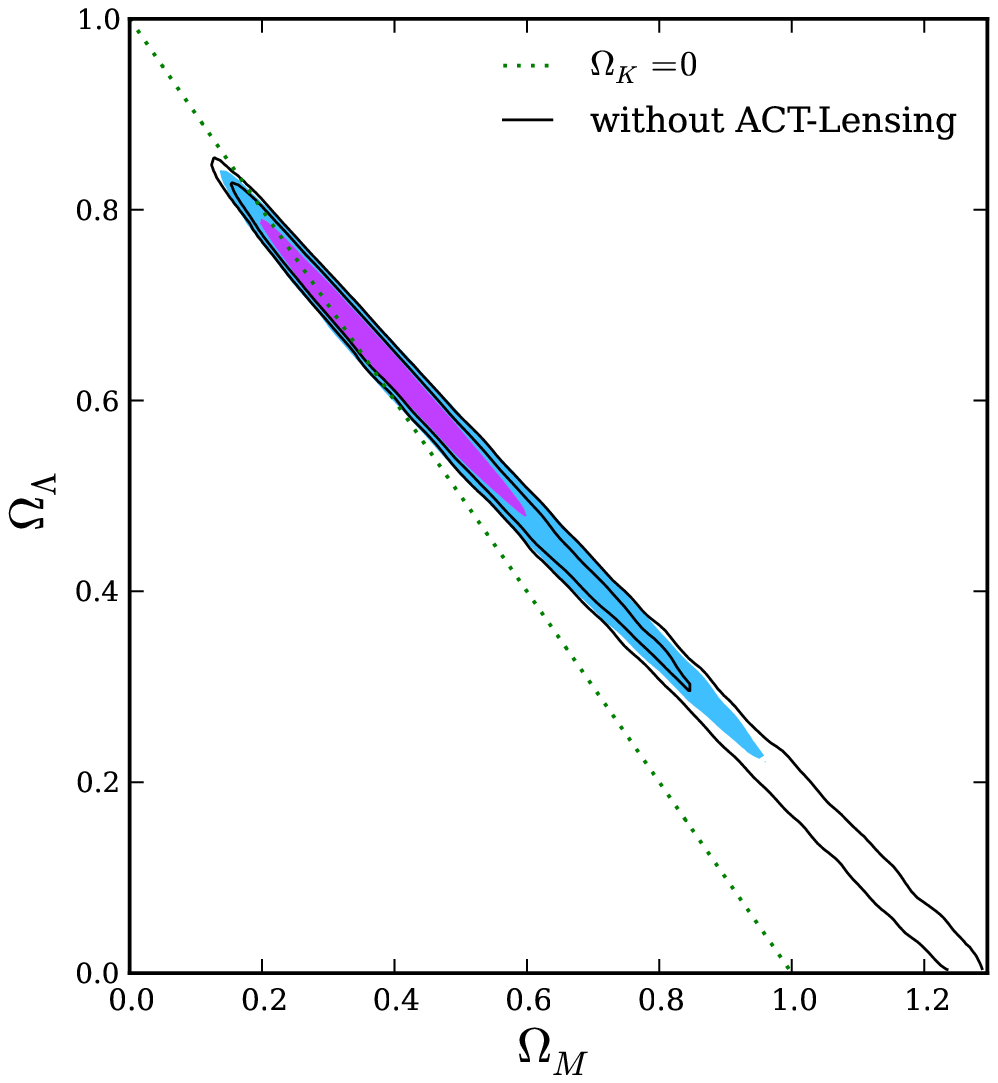}
\includegraphics[width=1.\columnwidth]{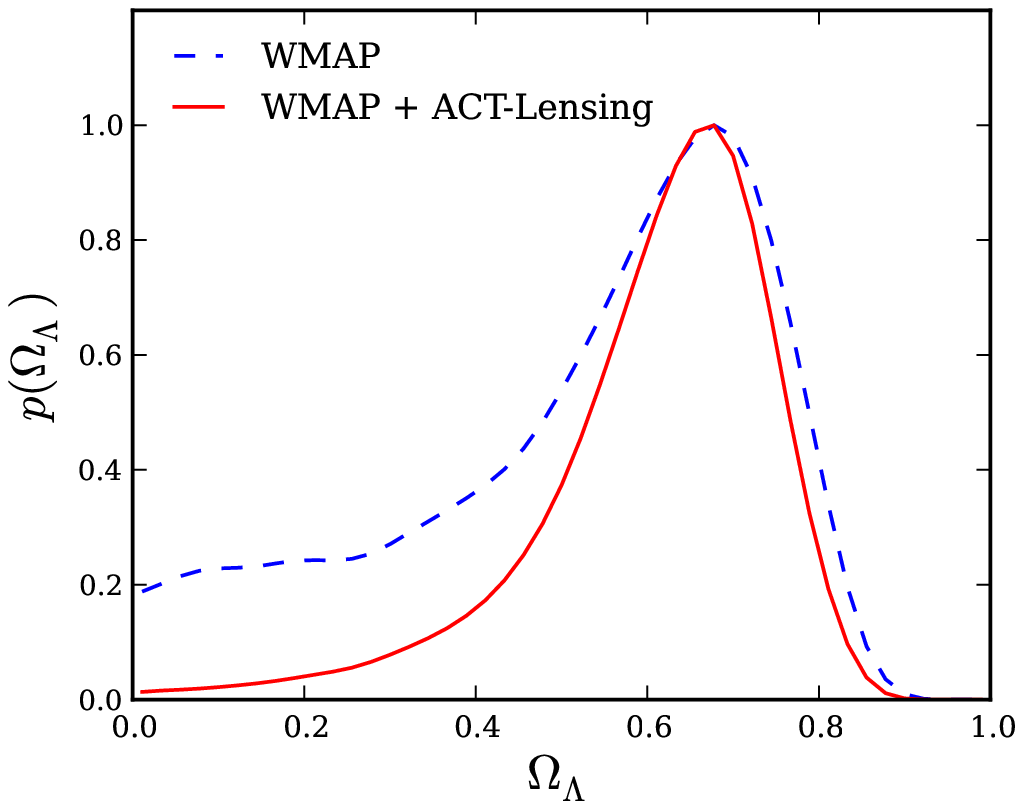}
\caption{Upper panel: Two-dimensional marginalized posterior probability for $\Omega_m$ and $\Omega_\Lambda$ (68\% and 95\% C.L.s shown). Colored contours are for WMAP + ACT Lensing, black lines are for WMAP only. Using WMAP data alone, universes with $\Omega_\Lambda=0$ lie within the 95\% C.L. The addition of lensing data breaks the degeneracy, favoring models with dark energy. \\
Lower panel: One-dimensional marginalized posterior probability for $\Omega_\Lambda$ (not normalized). An energy density of $\Omega_\Lambda\simeq0.7$ is preferred even from WMAP alone, but when lensing data are included, an $\Omega_\Lambda = 0$ universe is strongly disfavoured.
\label{fig:constraints}}
\end{figure}

The two-dimensional marginalized distribution for $\Omega_\Lambda$ and $\Omega_m=1-\Omega_K-\Omega_\Lambda$ is shown in Fig. \ref{fig:constraints}, with 68\% and 95\% confidence levels, indicating the effect of adding the ACT lensing data. 

The distribution for WMAP alone is limited by the ISW effect in both the TE and TT power, but is still unbounded at $\Omega_\Lambda=0$. It is truncated with the addition of the lensing data, resulting in a two-dimensional 95\% confidence level that excludes $\Omega_\Lambda=0$. The one-dimensional probability density for $\Omega_\Lambda$, also shown in Fig.~\ref{fig:constraints}, further demonstrates how the CMB lensing data reduce the low-$\Omega_\Lambda$ tail of the probability distribution and break the geometric degeneracy. A universe without dark energy would give too large a lensing signal to be consistent with the data. With lensing data, the new confidence intervals for $\Omega_\Lambda$ are $0.61^{+0.14}_{-0.06}$ at 1$\sigma$ (68\% C.L.), $0.61^{+0.23}_{-0.29}$ at 2$\sigma$ (95\% C.L.) and $0.61^{+0.25}_{-0.53}$ at 3$\sigma$ (99.7\% C.L.), favoring a model with dark energy. Comparing the likelihood value for the best-fit $\Lambda \mathrm{CDM}$ model with the likelihood for the best-fit $\Omega_\Lambda=0$ model, we find that $\Omega_\Lambda=0$ is disfavored at $3.2\sigma$ ($\Delta \chi^2 \approx 11$, of which $\Delta \chi^2\approx 5$ arises from the WMAP spectra, mainly due to differences in the TE and TT power spectra for $\ell < 10$). The parameters of the best-fit $\Lambda \mathrm{CDM}$ model are consistent with constraints from other datasets such as the WMAP+BAO+$\mathrm{H}_0$ constraints of \cite{komatsu/etal:2011}. The effect of massive neutrinos on the lensing spectrum is different from the effect of $\Omega_\Lambda$; neutrino masses within the current bounds can only modify the shape of the spectrum by $< 5\%$ \cite{lesgourgues/etal:2006}, whereas the reduction in $\Omega_\Lambda$ considered here increases the spectrum on all scales by a much larger amount. Our constraints on $\Omega_\Lambda$ do not apply to models with
non-power-law primordial power spectra \cite{hunt/sarkar:2008}, as
such models predict lensing deflection power which is currently
indistinguishable from $\Lambda \mathrm{CDM}$ for $\ell > 100$.

\section{Conclusions}
We find that a dark energy component $\Omega_\Lambda$ is required at a $3.2\sigma$ level from CMB data alone. This constraint is due to the inclusion of CMB lensing power spectrum data, which probe structure formation and geometry long after decoupling and so break the CMB geometric degeneracy. Our analysis provides the first demonstration of the ability of the CMB lensing power spectrum to constrain cosmological parameters. It provides a clean verification of other measurements of dark energy. In future work, our analysis can be easily extended to give constraints on more complex forms of dark energy with $w\neq-1$. With much more accurate measurements of CMB lensing expected from ACT, SPT \cite{carlstrom/etal:2009}, Planck \cite{perotto/etal:2010}, and upcoming polarization experiments including ACTPol \cite{niemack/etal:2010}, lensing reconstruction promises to further elucidate the properties of dark energy and dark matter \cite{galli/etal:2010}. 
\begin{acknowledgments}
This work was supported by the U.S.\ NSF through awards AST-0408698, PHY-0355328, AST-0707731 and PIRE-0507768, as well as by Princeton Univ., the Univ. of Pennsylvania,
FONDAP, Basal, Centre AIUC, RCUK Fellowship (JD),  NASA grant NNX08AH30G (SD, AH, TM), NSERC PGSD (ADH), the Rhodes Trust (RH),
NSF AST-0546035 and AST-0807790 (AK), NSF PFC grant PHY-0114422 (ES),  KICP Fellowship (ES), SLAC no. DE-AC3-76SF00515 (NS), ERC grant 259505 (JD), BCCP (SD), and the NSF GRFP (BDS, BLS). We thank  B.\ Berger, R.\ Escribano, T.\ Evans, D.\ Faber, P.\ Gallardo, A.\ Gomez, M.\ Gordon, D.\ Holtz, M.\ McLaren, W.\ Page, R.\ Plimpton, D.\ Sanchez, O.\ Stryzak, M.\ Uehara, and Astro-Norte for assistance with ACT. ACT operates in the Parque Astron\'{o}mico Atacama in northern Chile under the auspices of Programa de Astronom\'{i}a, a program of the Comisi\'{o}n Nacional de Investigaci\'{o}n Cient\'{i}fica y Tecnol\'{o}gica de Chile (CONICYT). We thank Alex van Engelen for discussions and thank Matias Zaldarriaga and Duncan Hanson for comments on the draft.
\end{acknowledgments}

\bibliography{act2}
\end{document}